# Influence of seed power on mode instabilities in high power fiber amplifiers

**Rumao Tao, Xiaolin Wang, * Pu Zhou,\*\* Dong Zhi, Lei Si, and Zejin Liu**
*College of Optoelectronic Science and Engineering, National University of Defense Technology, Changsha 410073, China*
*Author e-mail address: *chinawxllin@163.com; \*\*zhoupu203@163.com*

**Abstract:** We use a semi-analytical model of stimulated thermal Rayleigh scattering to present theoretical study of seed power on mode instability with the presence of photodarkening. The behavior of mode instabilities as a function of seed power has been investigated. The nonlinear dependence of the threshold with lower seed powers is ascribed to the influence of gain saturation while the reduction behavior of the threshold with higher seed powers is put down to the effects of photodarkening, which agrees well with the experimental results.

## 1. Introduction
Mode instabilities have been under intense investigation during the past few years, and several research groups have carried out much work to investigate the underlying physical mechanisms [1, 2]. So far a common agreement on the thermal origin of the mode instabilities has been achieved, while the required phase shift is still a subject of discussion. Essentially two different assumptions have been proposed in the literature, and models for both assumptions have been developed, which can be used to explain and predict the experiments [3-5]. However, the dependence of mode instabilities on the seed power, experimentally reported in Refs. [5] and [6], has not been explained or investigated in details from the perspective of the stimulated thermal Rayleigh scattering assumption. In this paper, based on a semi-analytical model of the stimulated thermal Rayleigh scattering, the behavior of threshold power and maximal coupling frequency as a function of seed power has also been investigated, and good agreement with experimental results has been achieved.

## 2. Theoretical study
In this section, we first study the threshold behavior as a function of seed power without considering the effects of photodarkening. To model mode instability thresholds and the maximal coupling frequency, we use our semi-analytical STRS model (detailed in [7, 8]). The co-pumped cladding pumped amplifiers are unbent, step-index fibers with uniform doping across the full core. We define threshold as the signal output or pump power at which $LP_{11}$ content reaches 5%. We assume that mode instabilities are induced by the quantum noise, which is a minimum seed level and yields the highest thresholds. Even though the simulations are exemplified for the particular case of the quantum noise induced mode instabilities, the qualitative behavior of the threshold will be similar for other cases [4, 9, 10]. The fiber parameters used are listed in Table 1. These parameters are typical of high power ytterbium doped amplifiers.

**Table 1. Parameters of Test Amplifier**

| | | | |
|---|---|---|---|
| $n_{clad}$ | 1.45 | $N_{Yb}$ | $3.2 \times 10^{25}$ m$^{-3}$ |
| $NA$ | 0.054 | $L$ | 1.5m |
| $\lambda_p$ | 976nm | $\sigma_p^a$ | $2.47 \times 10^{-24}$ m$^2$ |
| $\lambda_s$ | 1032nm | $\sigma_p^e$ | $2.44 \times 10^{-24}$ m$^2$ |
| $h_q$ | 5000 W/(m$^2$K) | $\sigma_s^a$ | $5.8 \times 10^{-27}$ m$^2$ |
| $\eta$ | $1.2 \times 10^{-5}$ K$^{-1}$ | $\sigma_s^e$ | $5.0 \times 10^{-25}$ m$^2$ |
| $\kappa$ | 1.38 W/(Km) | $\tau$ | 901 µs |
| $\rho C$ | $1.54 \times 10^6$ J/(Km$^3$) | | |

The pump power, signal power and extracted power at the mode instability threshold as a function of the seed power is shown in Figure 1. Here, the extracted power is defined as the output average power at the threshold minus the seed power. It can be seen from Fig. 1(a) that, for seed power lower than 100W, the threshold signal power increases nonlinearly with the seed power increasing. For higher seed power, the threshold signal power increases linearly with the seed power increasing. It can be seen that the nonlinear dependence of the threshold with seed powers, which is prominent for low seed power as shown in Fig. 1(b). This increase of threshold power is due to the influence of gain saturation enhancing while the nonlinear dependence of threshold power results from the varying speed of gain saturation enhancing. For lower seed power, the enhancing of the gain saturation is rapid while it tends to slow down for higher seed power, which is confirmed by the behavior of the maximal coupling frequency in Fig.

1(c). One can find from Fig. 1 (c) that the increased seed power tends to increase gain saturation and thus decrease the maximal coupling frequency [11]. For lower seed power, the enhancing of the gain saturation is rapid while it tends to slow down for higher seed power, which results in that the decrease of the maximal coupling frequency is sharp in low seed power range and that is slow during high seed power range.

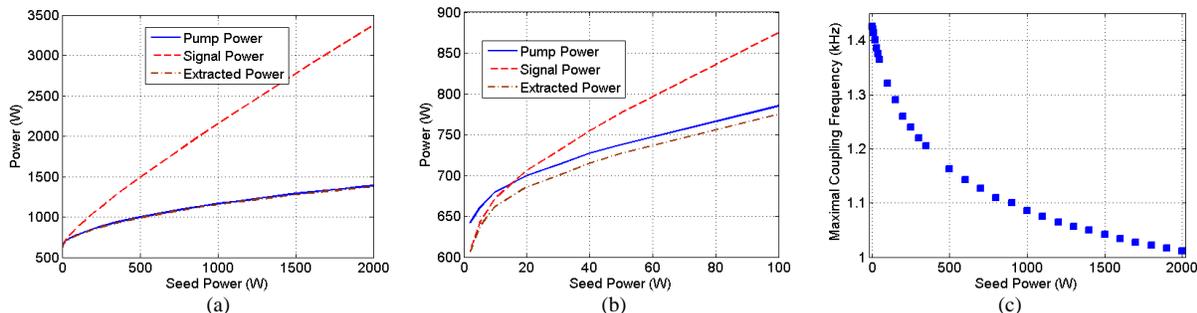

Fig. 1 Threshold power as a function of initial seed power (a), a detail description for the range from 0W to 100W (b), and the maximal coupling frequency as a function of seed power (c).

Then we studied the threshold behavior with photodarkening, which is shown in Fig. 2. The photodarkening mode employed in the calculation is similar to that in [12]. In [13], it shows that there is a nearly quadratic dependence of the equilibrium PD losses on the (local) population density in the upper laser level in a fiber, while a linear dependence has been concluded in [14]. So we calculated the equilibrium photodarkening loss distribution by multiplying the inversion level or the square of the inversion level with a value α, which is chosen to make the corresponding reduction in efficiency. In the calculations, the value of α corresponds to the reduction of efficiency was adjusted when seeded at the power of 10W, and then the same α was used for the whole seed power range. It can be seen that, no matter which formula was employed, the extracted power at the threshold increases nonlinearly as seed power increases until the maximal values were reached. Then the extracted power at the threshold reduced with higher seed powers after experiencing a region of saturation region, which agrees with the reported experimental observation in [5, 6]. The difference between the two formula lies in that severer photodarkening was required to reproduce the trend when the quadratic formula was employed, and the efficiency reduction of 20% was required. The aforementioned threshold behavior is a conjunction effect of gain saturation and photodarkening. For lower seed power, the increase of gain saturation is sharp, and the gain saturation increase is the dominant effects, which is why, when the seed power increases, the extracted power increases rapidly. For higher seed power, the increase of the gain saturation slows down, and the effect of photodarkening, which decreases the threshold, gradually compensates and then exceeds the effect of gain saturation increase, which results that the extracted power at the threshold gradually saturates and ultimately reduces.

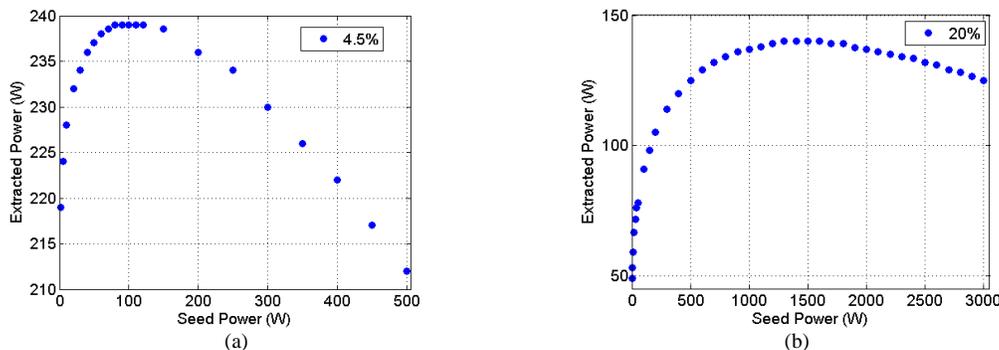

Fig. 2 Extracted power at the threshold as a function of initial seed power for 50/200 fiber when a linear formula was employed (a) and a quadratic formula was employed.

The maximal coupling frequency as a function of seed power was shown in Fig. 3. One can find that, under the same seed power, the maximal coupling frequency reduces for the case with photodarkening, which is due to that the photodarkening-induced absorption is strongest near the outer edge of the core, and this will tend to reduce the maximal coupling frequency [11]. Much work has been carried out by different research groups to investigate the physical mechanisms related to photodarkening, but the physical origin of it is still a subject of discussion. So it is difficult to accurately model the mode instabilities with the presence of photodarkening, but the results should give a reasonable indication of the influence of photodarkening.

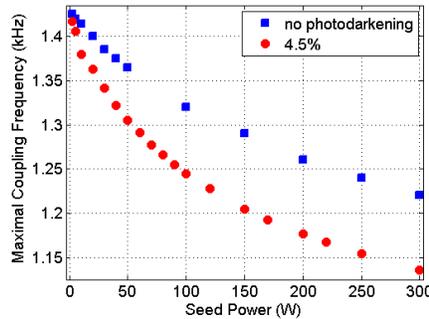

Fig. 3 The maximal coupling frequency as a function of initial seed power.

## 3. Conclusions

In summary, based on our semi-analytical model of the stimulated thermal Rayleigh scattering, we presented a theoretical study of mode instabilities in high power fiber lasers, and found that the nonlinear dependence of the threshold is irrespective of photodarkening while the reduction behavior shows up when photodarkening effects was taken into consideration. We found that the nonlinear behavior of extracted power was ascribed to gain saturation, and the reduction behavior was the associative effect of gain saturation and photodarkening. The calculated results agree with the recent reported experimental observation.